\documentclass[twocolumn,twoside,slac_two]{revtex4}
\usepackage{graphicx}
\usepackage{fancyhdr}
\newcommand{\lsi}{LS~I~+61$^{\circ}$303~}
\newcommand{\lsp}{LS~I~+61$^{\circ}$303}
\newcommand{\mnras}{MNRAS}
\newcommand{\aap}{A\&A}
\newcommand{\na}{New Astronomy}

\begin{document}

\title{Analysis of hard X-ray/high energy data from \lsi based on implications from its 4.6 yr periodicity}

%

\author{Zimmermann, L.}
\affiliation{Max Planck Institute for Radio Astronomy, Bonn, Germany}
\author{Grinberg, V.}
\affiliation{Remeis Observatory/ECAP/FAU, Bamberg, Germany}

\author{Massi, M.}
\affiliation{Max Planck Institute for Radio Astronomy, Bonn, Germany}
\author{Wilms, J.}
\affiliation{Remeis Observatory/ECAP/FAU, Bamberg, Germany}

\begin{abstract}
The most peculiar radio characteristics of the TeV emitting high-mass X-ray binary \lsi are two periodicities: A large periodic outburst which exhibits the same period as the orbit (phase $\Phi$) and a second periodicity of 1667 days (phase $\Theta$) which modulates the orbital phase and amplitude of the large outburst. Recent analysis of the radio spectral index present strong evidence for the presence of the critical transition from optically thick emission (related to a steady jet) to an optically thin outburst (related to a transient jet) as in other microquasars. In parallel, a switch from a low/hard X-ray state to a transitional state (e.g. steep power law state) would be expected. We show how the critical transition from optically thick emission to an optically thin outburst is modulated by $\Theta$. Folding over a too large $\Theta$ interval mixes up important information about the outbursts and can yield a false picture of the emission behaviour of the source along the orbit. We therefore analyse the implications of this long period for treatment of hard X-ray/high energy data obtained from \lsi, e.g. with \textit{Fermi}-LAT or INTEGRAL, taking into account this long-term periodicity.

\end{abstract}

\maketitle

\thispagestyle{fancy}


\section{Introduction}
\lsi is an X-ray binary formed by a compact object and a massive star with
an optical spectrum typical for a rapidly rotating B0 V star
\cite{Hutchings}. The nature of the compact object, if neutron star or black hole, is still unknown due to the large uncertainty in the inclination of the object \cite{Aragona,Casares}. It travels around its companion star on an eccentric orbit with a period of 26.5 days \cite{Gregory02}.

Radio spectral index analysis by \cite{MassiKaufman} have found two peaks along the orbit of \lsp. Each peak shows the microquasar characteristic of a switch from a steady (optically thick) to a transient (optically thin) jet. Furthermore, each peak in the radio spectral index is accompanied by two distinguishable peaks in the radio flux. This confirms that there are really two different outbursts (an optically thick and an optically thin one). Also high energy observations with EGRET \cite{Massi05} and \textit{Fermi}-LAT  \cite{Abdo} indicate two peaks along the orbit. The high energy peaks are supposed to be due to inverse Compton upscattering of the UV photons from the donor star by the relativistic electrons of the jet, which in turn strongly attenuates the radio peak around periastron (see discussion in \cite{MassiZimmermann} and \cite{Massi10}).

 \begin{figure}
   \centering
\includegraphics[width=.3\textheight]{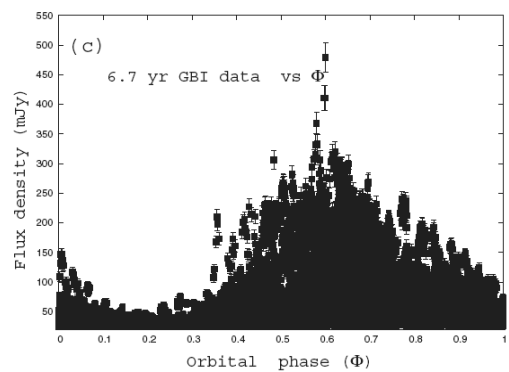}
      \caption{Radio light curve of \lsi from the Green Bank
interferometer vs. orbital phase \cite{MassiKaufman}. The large
outburst towards apastron is clearly observable. The broadness of the
curve stems from the 4.6 yr period which modulates the orbital occurrence
of the peak.}
         \label{Fig1}
   \end{figure}
It has been shown from theory that for a microquasar with an eccentric
orbit (\lsi: e=0.54-0.7 \cite{Aragona,Casares}), indeed, the different
relationship between the accretion rate for density and velocity described
by \cite{Bondi}, creates two peaks in the accretion rate curve, one at
periastron and a second one towards apastron \cite{Taylor92,MartiParedes,BoschRamon,Romero2007}.

The clear periodicity in \lsi is amongst its most peculiar
characteristics. The orbital period ($\Phi$) modulates the flux at radio,
H-alpha, X-rays and also in gamma-rays (high and very high energy)
\cite{Gregory02,LiuYan05,Grundstrom07,Smith09,Chernyakova06,Zhang10,Massi05,Abdo,Acciari09,Albert09}.
But \lsi holds another peculiarity: A second period, which modulates the
radio flux over a long period of $\Theta$=4.6 yr \cite{Gregory02}. This
modulation in radio is shown in Fig. \ref{Fig1} (radio flux density vs. $\Phi$), which shows how the peak flux shifts around apastron, and in Fig \ref{Fig3}, where the radio flux density is given vs. $\Theta$. In addition, this long period has also been shown in H-alpha (see Fig. 2 in \cite{Zamanov}). It is suggested that this modulation could be due to periodic shell ejections from the circumstellar disk of the Be star \cite{GregoryNeish}.

\section{High energy observations and the radio spectral index}
\begin{figure}
   \centering
\includegraphics[width=.3\textheight]{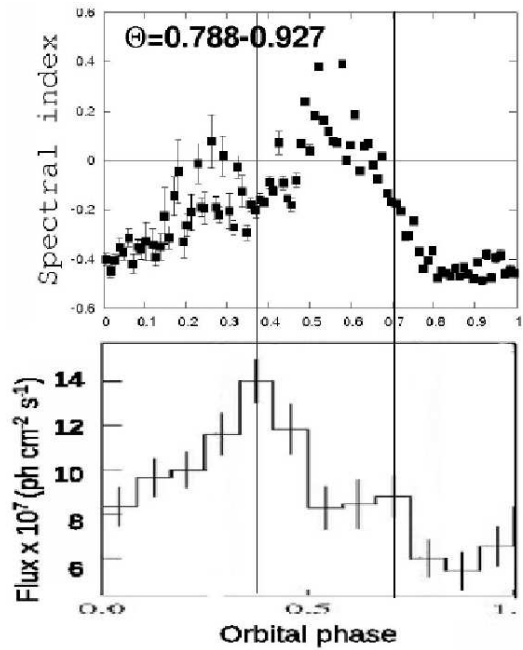}
      \caption{Top: Radio spectral index data for $\Theta$=0.788-0.927
from GBI \cite{Massi10}. Bottom: \textit{Fermi}-LAT lightcurve obtained by \cite{Abdo}. The two vertical lines indicate the peaks around periastron and around apoastron. They both correspond to a negative spectral index in the radio data, which corresponds to optically thin emission in both cases.}
         \label{Fig4}
   \end{figure}
\begin{figure}
   \centering
\includegraphics[angle=-90, width=.3\textheight]{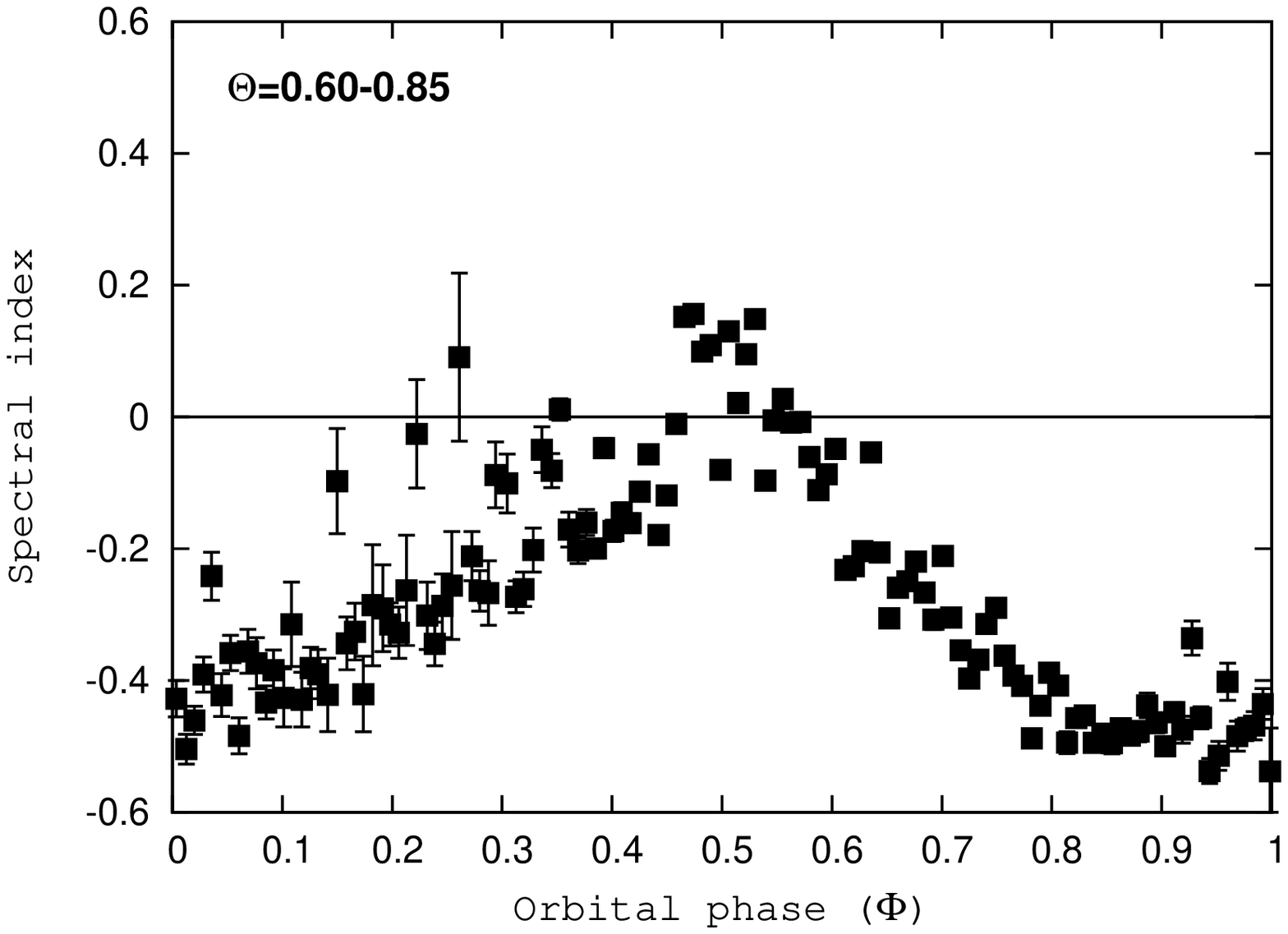}
\includegraphics[angle=-90, width=.3\textheight]{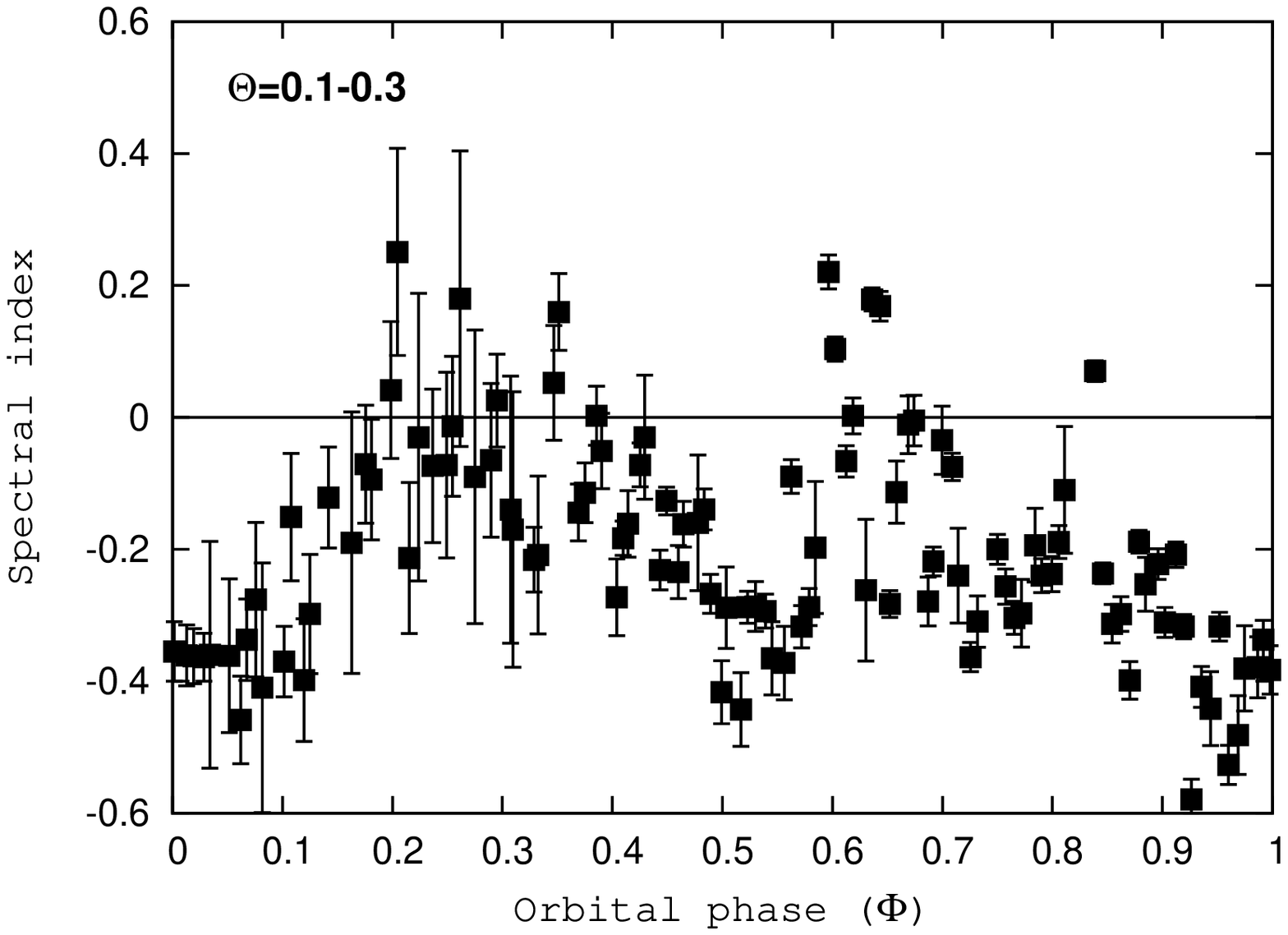}
      \caption{Top: Radio spectral index data for $\Theta$=0.6-0.85
from GBI \cite{Zimmermann}. Bottom: Radio spectral index data for $\Theta$=0.1-0.3 from GBI \cite{Zimmermann}.}
         \label{Fig5}
   \end{figure}

The radio spectral index analysis by \cite{MassiKaufman} have shown the
importance of the large phase $\Theta$ for the analysis of radio data from
\lsi. It has thereby proven that the radio outbursts really consist of two
consecutive outbursts. These outbursts have completely different
characteristics. In the microquasar model, an optically thick outburst in
radio is associated with a steady jet, centered on the compact object. The
optically thin outburst comes from a transient jet, detached from the
central engine. In the unified model of X-ray states with radio jets, the
radio states in microquasars are clearly associated with two X-ray states.
A steady jet corresponds to the low hard state, where the X-ray spectrum
is characterized by $\Gamma\approx$1.5, whereas a transient jet
corresponds to a transitional state, e.g. the steep power law state with $\Gamma>$2.4 \cite{FenderBelloniGallo,McClintockRemillard04,McClintockRemillard06}.
Furthermore, high energy and very high energy emission are directly
connected to the transitional state (steep power law state) as the power law is without cut-off and extends into the gamma-ray regime. As a matter of fact, when \lsi was detected by MAGIC and VERITAS the spectrum was always fitted with a power law with an index $\Gamma \geq$2.4 \cite{Albert09,Acciari09,Acciari2011,Jogler2011}. Moreover, as discussed in the next section, INTEGRAL observations seem to indicate a change of the photon index consistent with a change from the low hard to the transitional state in agreement with the radio spectral index.

If the radio spectral index now tells us about the nature of the outburst,
then the high and very high energy observations should corroborate this
nature and might give additional information about the emission processes.
But in order to get the complete information from the radio spectral index
for \lsi, it was neccessary to take into account the long period $\Theta$.
Therefore, to compare high energy data with the radio spectral index, it
is neccessary as well to take only data of the same $\Theta$ phase. \cite{Massi10} has done so e.g. for the first eight months of \textit{Fermi}-LAT observations and the results (see Fig. \ref{Fig4}) show that in this comparison the high energy peaks established by \textit{Fermi}-LAT both correspond to the optically thin outbursts of \lsp. One could then draw the conclusion that the population of relativistic particles, producing the optically thin outburst (and
therefore the transient jet), are also responsible for the production of
the high energy emission. Furthermore, in Fig. \ref{Fig5} are shown radio spectral index curves vs. the orbital phase for two different $\Theta$ intervals. In particular the orbital occurrence of the second peak is different in the different intervals. It is evident that folding data only on the orbital period, without respect to the long period, might result in mixing up different ejection processes, because the orbital phase of the peaks and therefore of the
switch from optically thick to thin emission does not stay the same over
the 4.6 yr period. This point will be discussed in the next section.

\section{INTEGRAL observations of \lsi and $\Theta$}

\begin{figure}
   \centering
\includegraphics[width=.35\textheight]{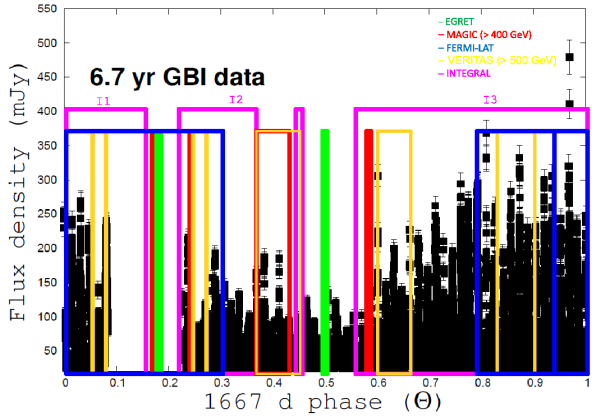}
      \caption{Published high energy observations of \lsi by EGRET, MAGIC,
\textit{Fermi}-LAT, VERITAS and INTEGRAL shown in the context of the 4.6 yr radio period ($\Theta$) \cite{Zimmermann}. The underlying light curve gives the radio flux at 8.3
GHz against $\Theta$ \cite{MassiKaufman}. The minimum
lies around $\Theta\approx$0.5 and the maximum around $\approx$0.97. The
three largest $\Theta$ intervals where the source was observed with
INTEGRAL are named here I1, I2 and I3.}
         \label{Fig3}
   \end{figure}

In Fig. \ref{Fig3}, high energy and very high energy observations of \lsi
with different instruments are shown with respect to the $\Theta$
intervals in which they have been carried out. Both, MAGIC and VERITAS probe the minimum and the maximum parts of $\Theta$ \cite{Albert09,Acciari2011,Jogler2011}. Due to the long integration times for the data, the sampling within these intervals is not very strong, although EGRET observed one complete orbital cycle at each of the two given intervals \cite{Albert09,Acciari09,Massi05}. The latest published coverage by \textit{Fermi} is of 2.5 years \cite{Hadasch2011}.

For INTEGRAL, the long-term monitoring of \lsi covers vast parts of the
4.6 yr period. We have denoted the three big INTEGRAL observations with
respect to $\Theta$ by I1, I2 and I3, as noted in Fig. \ref{Fig3}. Folding the data by orbital phase, of course, increases the
sampling. By doing so, it was established that the emission between 10-100
keV is clearly modulated by the orbital phase
\cite{Chernyakova06,HermsenKuiper07,Zhang10}.

Following the radio spectral index though (see Fig. \ref{Fig5}), folding over almost a complete $\Theta$ cycle would imply that even though the resulting light curves could show the overall periodicity of the source at these energies, a deeper insight about the ejections can be obscured by mixing different ejection processes (namely optically thick and thin ejections). The spectral analyses can then get corrupted, because of the same mixing of ejection processes.

I1 and I3 are covering most of the maximum of $\Theta$ (see Fig. \ref{Fig3}). An interpretation of these data with respect to the flux and the spectrum should be less corrupted than a mix of data from the maximum and the minimum. In fact, \cite{Chernyakova06}, who used most of I3 and parts of I1, have found not only the modulation of the lightcurve with the orbital period, but also found that along the orbit the spectral index $\Gamma$ changed
from $\approx$1.5 around periastron to $\approx$3.2 around apastron. This
result strengthens the two-peak accretion model, as for the
$\Theta$-intervals covered by these observations an optically thin
ejection (a transient jet) is expected around apastron
(see \cite{MassiKaufman} and discussions therein) As mentioned above, in the
unified X-ray states model with radio jets, a transient is associated with a transitional state, e.g. the steep power law, characterized by a spectral index $\Gamma>$2.4
\cite{FenderBelloniGallo,McClintockRemillard04,McClintockRemillard06}.

\section{Conclusions and Discussion}

In the high mass X-ray binary \lsp, two clear radio periodicities are
present, one coincident with its orbital period (see Fig. \ref{Fig1}) and
the other modulating the strength of the large radio outburst over a
period of 4.6 yr \cite{Gregory02,GregoryNeish}. Both periods have also been
observed in H-alpha emission. This system is amongst a few to have been
detected not only in radio and X-rays, but also at high and very high
energies. It is from these observations that, together with radio spectral
index analysis, important insights into the nature of the system and its
emission processes can be deduced. The radio spectral index tells us about
the nature of the observed outburst. There are two different kinds of
outburst: optically thick (spectral index $\alpha>$ 0) and optically thin
($\alpha<$ 0). The first is attributed in the microquasar model to a slow
outflow, while the latter is associated with an ultrarelativistic
transient jet.

Recent spectral index analysis of \lsi show that a subsequent realization
of these two types of outburst takes places not only once, but twice
along the orbit (see Fig. 2 top). This can be explained by the two peak
accretion/ejection microquasar model for systems with an eccentric orbit
like \lsi (e=0.54-0.7) (\cite{MassiKaufman} and references within). In
this model an expected radio peak around periastron is attenuated by
inverse Compton losses due to the dense stellar UV field. In the first eight months observations with \textit{Fermi}-LAT, a related peak at high energies is observed (see Fig. 2 at $\Phi$=0.3-0.4). A second (but smaller) high energy peak is observed coincident with the large radio outburst around apastron ($\Phi$=0.65-0.75). Furthermore, as discussed in \cite{MassiKaufman}, the two peak shape of the $\alpha$ vs. $\Phi$ curve varies with $\Theta$, as does the distance of the two peaks (see also Fig. 2 here). The two-peak microquasar model of \cite{MartiParedes} predicts these variations in the accretion curve by incorporating changing wind velocities for the Be star (see their Fig.6). This behaviour should be seen at other wavelengths as well, due to the connected emission processes. In fact, after the first eight months, \textit{Fermi}-LAT  detected an increase in the overall flux level ($\Theta\approx$ 0.92) and a broadening of the peak shape \cite{Hadasch2011}. These variations are consistent with the observation that $\alpha$ varies with $\Theta$. Variations are also seen with VERITAS and MAGIC. \lsi was detected around apastron until 2008 and then became quiescent between 2008-2010, where no detection was reported by VERITAS, while MAGIC reported only weak detection around apastron. Then in October 2010, VERITAS detected the source again, this time around periastron \cite{Albert09,Acciari2011,Jogler2011}. 

The strong connection between radio and HE/VHE emission underlines therefore the importance of the radio periods for the analysis of e.g. INTEGRAL data and can and should in principle be extended to other high energy instruments.

\bigskip 

\begin{acknowledgments}
The work of L. Zimmermann is partly supported by the German Excellence
Initiative via the Bonn Cologne Graduate School.
\end{acknowledgments}

\bigskip

\end{document}